\documentclass[12pt]{article}

\title{Whitham-Broer-Kaup Systems in Multi-dimensions: Quantum and Resonant NLS Connections}

\author{Oktay K Pashaev$^1$ and Colin Rogers$^2$\\
${^1}$Department of Mathematics, Izmir Institute of Technology,\\ Izmir 35430, T\"urkiye 
\\
${^2}$School of Mathematics and Statistics, University of New South Wales,\\ Sydney NSW 2052, Australia}

\begin{document}

\maketitle              % typeset the title of the contribution

\begin{abstract}
  An overview is presented of quantum and resonant nonlinear Schr\"odinger equation links to Whitham-Broer-Kaup type systems.
A novel $n+1$ dimensional extension of the Whitham-Broer-Kaup hydrodynamic system is constructed with connection to an equivalent 
multi-dimensional resonant NLS equation. Hybrid Ermakov-Painlev\'e II and associated Painlev\'e  XXXIV integrable similarity reductions 
are derived.
\end{abstract}

%\footnote{An example footnote.}

 %
\section{Introduction}
The $1+1$-dimensional Whitham-Broer-Kaup (WBK) system had its origin in work by Whitham \cite{W} on a variational approach to nonlinear wave theory. 
It was subsequently derived independently by Broer \cite{Broer} as a Hamiltonian treatment of long wave propagation and by Kaup \cite{Kaup} in connection
with a nonlinear wave equation amenable   the inverse scattering transform of modern soliton theory. In \cite{RogersPashaev}, a novel $2+1$-dimensional extension 
of the Whitham-Broer-Kaup system was constructed, which is equivalent to a resonant solitonic Davey-Stewartson equation as originally introduced in a capillarity
context in \cite{RogersYipChow} and which incorporates a de Broglie-Bohm quantum potential. The $\bar\partial$-dressing method for such integrable $2+1$-dimensional system 
has been subsequently investigated in detail by Nabelek and Zakharov \cite{NabelekZakharov}.

Resonant nonlinear Schr\"odinger equations incorporating a de Broglie-Bohm potential have origin in Madelung hydrodynamics \cite{ModPhysLettA}. In \cite{PashaevLeeRogers},
their potential admittance, subject to a super-critical parameter constraint, of novel solitonic fission and fusion phenomena was established for a subclass of nonlinear Schr\"odinger type 
equations with underlying  Hamiltonian structure as described  in \cite{MalomedStenflo}.

The resonant NLS equation has subsequently proved to have diverse physical applications, notably in capillarity theory \cite{RogersSchief1}, \cite{Rogers1}, cold plasma physics \cite{LeePashaevRogersSchief} and nonlinear 
optics \cite{Kudryashov}.

Solitonic resonance phenomena in Whitham-Broer-Kaup systems have previously been analyzed in \cite{P1}, \cite{Lee}. Here, "inter alia" a diversity of properties of quantum hydrodynamic systems
is detailed together with resonant nonlinear Schr\"odinger equation connections. A multi-dimensional extension of a symmetry representation previously applied in a nonlinear optical context 
in \cite{Giannini} is applied to the $n+1$-dimensional resonant nonlinear Schr\"odinger equation subject to a super-critical parameter condition applied to the de Broglie-Bohm term.
Reduction is thereby obtained to a canonical hybrid Ermakov-Painlev\'e II equation of a type with origin in wave packet reductions admitted by multi-dimensional resonant Manakov-type
vector NLS systems \cite{Rogers2}. Ermakov systems with genesis in the classical work of \cite{Ermakov} have proved to possess important physical applications notably in nonlinear optics 
\cite{RogersMalomed1}, \cite{RogersMalomed2} and other diverse areas of physics and continuum mechanics \cite{RogersSchief2}. Hybrid integrable Ermakov-Painlev\'e II reduction since its 
introduction in \cite{Rogers2} has been subsequently applied in Korteweg capillarity theory \cite{RogersClarkson1}, cold plasma physics \cite{RogersClarkson2} and in the analysis of nonlinear
boundary problems for the Nernst -Planck electrolytic systems \cite{ARogers}. In the present work, a novel link between the $n+1$-dimensional resonant super-critical NLS equation and a multi-dimensional
extension of the Whitham-Broer-Kaup system is applied to derive Ermakov-Painlev\'e II and associated integrable Painlev\'e XXXIV symmetry reductions.  

\section{The Madelung transformation: background}
The non-stationary Schr\"odinger equation in quantum mechanics
\begin{equation}
i\hbar \Psi_t + \frac{\hbar^2}{2m} \nabla^2 \Psi - U \Psi = 0 \label{Schrodinger1}
\end{equation}
 was transformed by Madelung to  the quantum fluid representation
\begin{eqnarray}
m( {\bf v}_t + ({\bf v}\cdot \nabla){\bf v}) &=& - \nabla \left( U + \frac{\hbar^2}{2m} \frac{\nabla^2 \sqrt{\rho}}{\sqrt{\rho}} \right), \label{M1}\\
\rho_t + \nabla (\rho {\bf v})&=&0\label{M2}
\end{eqnarray}
wherein $\rho = |\Psi|^2$ and 
\begin{equation}{\bf v}_q = \frac{\hbar}{2 m} \nabla \ln \rho \end{equation} 
is the "quantum" or osmotic (stochastic) velocity.
The latter represents the real part of the complex velocity 
\begin{equation}
{\bf V} = \frac{\hbar}{m} \frac{\nabla \Psi}{\Psi} = {\bf v}_q + i {\bf v}_c,\label{complexvelocity}
\end{equation}
the imaginary part of which constitutes the electron velocity, or the Madelung-Landau-London local mean velocity (current velocity)
\begin{equation}
{\bf v}_c = \frac{i\hbar}{2m \rho} (\Psi \nabla \bar\Psi - \bar\Psi \nabla \Psi) = \frac{\hbar}{m} \Im \frac{\nabla \Psi}{\Psi} = \frac{\nabla S}{m}.
\end{equation}
The Schr\"odinger equation (\ref{Schrodinger1}) in terms of the complex velocity (\ref{complexvelocity}) yields
\begin{equation}
i {\bf V}_t + ({\bf V}\cdot \nabla){\bf V} + \frac{\hbar}{2m} \nabla^2 {\bf V} = \frac{1}{m} \nabla U
\end{equation}
whence the Madelung transformation (\ref{complexvelocity}) then plays the role of a complex version of the classical Cole-Hopf transformation.

\subsection{Quantum WBK fluid representation}

Here drift velocities, namely, mean forward (backward) velocities are introduced  as combinations of ${\bf v}_c$ and ${\bf v}_q$ according to
\begin{eqnarray}
{\bf v}^+ = \frac{1}{2} ({\bf v}_q + {\bf v}_c) = \frac{\hbar}{m} \nabla R + \frac{1}{m} \nabla S, \\
 {\bf v}^- = \frac{1}{2} ({\bf v}_q - {\bf v}_c) = \frac{\hbar}{m} \nabla R - \frac{1}{m} \nabla S,
\end{eqnarray}
while the wave function is represented as
\begin{equation}
\Psi = e^{R + \frac{i}{\hbar} S}.
\end{equation}
The classical Schr\"odinger equation (\ref{Schrodinger1}) is then equivalent to the system
\begin{eqnarray}
R_t + \frac{1}{2m} (\nabla^2 S + 2 \nabla R \nabla S) =0, \\
S_t + \frac{1}{2m} (\nabla S)^2 + U = \frac{\hbar^2}{2m} [\nabla^2 R + (\nabla R)^2].
\end{eqnarray}
The preceding produces a  continuity equation
\begin{equation}
\rho_t + \nabla (\rho {\bf v}_c) =0,
\end{equation}
together with a pair of coupled hydrodynamic equations
\begin{eqnarray}
{{\bf v}_c}_t + \frac{1}{2} \nabla ({\bf v}_c)^2 + \frac{1}{m} \nabla U = \nabla \left[\frac{\hbar}{2m} (\nabla \cdot {\bf v}_q) + \frac{1}{2} ({\bf v}_q)^2 \right], \\
{{\bf v}_q}_t + \nabla \left[\frac{\hbar}{2m} (\nabla \cdot {\bf v}_c) + ({\bf v}_c \cdot {\bf v}_q)\right] = 0.
\end{eqnarray}
These can be re-written as equations for drift velocities ${\bf v}^+$ and ${\bf v}^-$, with
\begin{equation}
{\bf v}^+ + {\bf v}^- = \frac{\hbar}{2 m} \frac{\nabla \rho}{\rho}
\end{equation}
whence
\begin{eqnarray}
m {\bf v}^+_t + \frac{\hbar}{2} \nabla^2 {\bf v^+} &=& \nabla \left[- m ({\bf v}^+)^2 - \frac{1}{2} U + \frac{\hbar^2}{2m} \frac{\nabla^2 \sqrt{\rho}}{\sqrt{\rho}}\right], \\
-\rho_t + \frac{\hbar}{2m} \nabla^2 \rho &=& 2 \nabla (\rho {\bf v}^+),
\end{eqnarray}
where $rot\, {\bf v}^+ =0$ together with
\begin{eqnarray}
-m {\bf v}^-_t + \frac{\hbar}{2} \nabla^2 {\bf v^-} &=& \nabla \left[- m ({\bf v}^-)^2 - \frac{1}{2} U + \frac{\hbar^2}{2m} \frac{\nabla^2 \sqrt{\rho}}{\sqrt{\rho}}\right], \\
\rho_t + \frac{\hbar}{2m} \nabla^2 \rho &=& 2 \nabla (\rho {\bf v}^-),
\end{eqnarray}
where $rot\, {\bf v}^- =0$. These systems can be considered as independent and admit symmetry transformation: ${\bf v}^+ \longleftrightarrow {\bf v}^-$, $t \longleftrightarrow -t$.
They represent hydrodynamic equations with drift velocities   ${\bf v^+}$ and ${\bf v^+}$  in turn.

\subsubsection{Classical limit}

In the classical limit, $\hbar = 0$, the system takes form
\begin{eqnarray}
{\bf v}^+_t + \nabla ({\bf v}^+)^2 = - \frac{1}{2m} \nabla  U, \label{Euler}\\
\rho_t + 2 \nabla (\rho {\bf v}^+) =0, 
 \end{eqnarray}
where (\ref{Euler}) is the Euler equation (the dispersionless Burgers equation) 
\begin{equation}
{\bf v}^+_t + 2({\bf v}^+ \cdot \nabla) {\bf v}^+ = - \frac{1}{2m} \nabla  U.
\end{equation}

\subsubsection{Semi-classical limit}

In the next approximation to the $1^{st}$ order in $\hbar$, the quantum potential term is neglected resulting in the system
\begin{eqnarray}
 {\bf v}^+_t + \frac{\hbar}{2} \nabla^2 {\bf v^+} + \nabla ({\bf v}^+)^2 &=& - \frac{1}{2m} \nabla U, \\
-\rho_t + \frac{\hbar}{2m} \nabla^2 \rho &=& 2 \nabla (\rho {\bf v}^+). 
\end{eqnarray}

\subsection{Nonlocal $n+1$ dimensional NLS}

If in the Schr\"odinger equation (\ref{Schrodinger1}) the potential $U$ is
\begin{equation}
U({\bf x}, \xi, t) = \nu \int^1_0 |\Psi({\bf x}, \xi, t)|^2 d\xi,
 \end{equation}
where the wave function depends on the extra parameter $\xi$, the equation
\begin{equation}
i\hbar \Psi_t + \frac{\hbar^2}{2m} \nabla^2 \Psi - \nu \left(\int^1_0 |\Psi({\bf x}, \xi, t)|^2 d\xi\right)
 \Psi = 0, \label{Schrodinger2}
\end{equation}
results.
The infinite system of NLS equations in $n+1$ dimensions with non-local self-interaction for $n=1$ was considered in \cite{Zakharov}.
For a discrete set of parameter $\xi$, as $\xi_1, \xi_2,...,\xi_N$, we have $N$ wave functions $\Psi^{(k)} = \Psi({\bf x}, \xi_k, t))$, $k=1,2,...,N$, satisfying 
the system of vector $U(N)$  NLS equations
\begin{equation}
i\hbar \Psi^{(k)}_t + \frac{\hbar^2}{2m} \nabla^2 \Psi^{(k)} - \nu \left(\sum^N_{n=1} |\Psi^{(n)}({\bf x}, t)|^2 \right)
 \Psi^{(k)} = 0. \label{Schrodinger3}
\end{equation}
In the limit $N \rightarrow \infty$, it gives the infinite component vector $U(\infty)$ NLS. If $\Psi({\bf x}, \xi, t)$ is continuous in variable $\xi$, and domain of integration 
is the whole line $-\infty < \xi < \infty$, the equation can be considered as nonlocal NLS in $n+1$ dimensions. In the particular case $n=1$ the resulting two-dimensional equation
was solved by the bilinear Hirota method. Arbitrary  $M$-soliton solutions localized in the plane were derived in\cite{Maruno}.  

The hydrodynamic system for the forward drift velocity in this case is
\begin{eqnarray}
m {\bf v}^+_t + \frac{\hbar}{2} \nabla^2 {\bf v^+} &=& \nabla \left[- m ({\bf v}^+)^2 - \frac{1}{2} U + \frac{\hbar^2}{2m} \frac{\nabla^2 \sqrt{\rho}}{\sqrt{\rho}}\right], \\
-\rho_t + \frac{\hbar}{2m} \nabla^2 \rho &=& 2 \nabla (\rho {\bf v}^+), \\
U({\bf x}, \xi, t) &=& \nu \int^1_0 \rho({\bf x}, \xi, t) d\xi.
\end{eqnarray}

\subsubsection{Classical limit}

In the classical limit $\hbar =0$ the preceding system becomes
\begin{eqnarray}
 {\bf v}^+_t + 2({\bf v}^+ \cdot \nabla) {\bf v}^+   + \frac{1}{2 m}\nabla  U &=&0, \\
\rho_t + 2 \nabla (\rho {\bf v}^+) &=& 0, \\
U({\bf x}, \xi, t) &=& \nu \int^1_0 \rho({\bf x}, \xi, t) d\xi.
\end{eqnarray}
This constitutes a multidimensional generalization of the Benney system for long shallow water waves. For n=1 the horizontal velocity $v^+(x,\xi,t)$ is a function of the layers in the 
z-direction, enumerated by the parameter $\xi$, $0 < \xi < 1$, while the surface geometry of the fluid is determined by  U(x,t). 

\subsubsection{Semi-classical limit}

In the semi-classical limit 

\begin{eqnarray}
m {\bf v}^+_t + \frac{\hbar}{2} \nabla^2 {\bf v^+} &=& \nabla \left[- m ({\bf v}^+)^2 - \frac{1}{2} U\right], \\
-\rho_t + \frac{\hbar}{2m} \nabla^2 \rho &=& 2 \nabla (\rho {\bf v}^+), \\
U({\bf x}, \xi, t) &=& \nu \int^1_0 \rho({\bf x}, \xi, t) d\xi
\end{eqnarray}
namely a Whitham-Broer-Kaup type system in multidimensions. For $n=1$ such a system was analyzed in \cite{Lee}.

\subsection{Nonlinear Schrodinger equation}

 If  $U = - \nu |\Psi|^2$, then the NLS equation becomes

\begin{equation}
i \hbar \Psi_t + \frac{\hbar^2}{2m} \nabla^2 \Psi + \nu |\Psi|^2 \Psi =0.
\end{equation}
This equation describes Bose gas evolving in $n$-space dimensions with delta pair interaction between particles and the phenomenological Ginzburg-Landau 
equation for the wave function of superfluid with Bose-Einstein condensate.

The corresponding hydrodynamic systems for drift velocities ${\bf v}^+$ and ${\bf v}^-$ are
\begin{eqnarray}
m {\bf v}^+_t + \frac{\hbar}{2} \nabla^2 {\bf v^+} &=& \nabla \left[- m ({\bf v}^+)^2 + \frac{\nu}{2} \rho + \frac{\hbar^2}{2m} \frac{\nabla^2 \sqrt{\rho}}{\sqrt{\rho}}\right], \label{drift1}\\
-\rho_t + \frac{\hbar}{2m} \nabla^2 \rho &=& 2 \nabla (\rho {\bf v}^+), \label{drift2}
\end{eqnarray}
and
\begin{eqnarray}
-m {\bf v}^-_t + \frac{\hbar}{2} \nabla^2 {\bf v^-} &=& \nabla \left[- m ({\bf v}^-)^2 + \frac{\nu}{2} \rho + \frac{\hbar^2}{2m} \frac{\nabla^2 \sqrt{\rho}}{\sqrt{\rho}}\right], \\
\rho_t + \frac{\hbar}{2m} \nabla^2 \rho &=& 2 \nabla (\rho {\bf v}^-),
\end{eqnarray}
respectively.
In the semi-classical  limit (\ref{drift1}), (\ref{drift2}) reduces to a  Whitham-Broer-Kaup system, namely
\begin{eqnarray}
m {\bf v}^+_t + \frac{\hbar}{2} \nabla^2 {\bf v^+} &=& \nabla \left[- m ({\bf v}^+)^2 + \frac{\nu}{2} \rho \right], \\
-\rho_t + \frac{\hbar}{2m} \nabla^2 \rho &=& 2 \nabla (\rho {\bf v}^+). 
\end{eqnarray}

\section{The resonant NLS equation in n-space dimensions }

The Bose gas with delta pair interaction between particles, under the  influence of a quantum potential is described by the resonant NLS equation 
($\hbar =1$, m=1/2) 
\begin{equation}
i\Psi_t + \nabla^2 \Psi + \nu |\Psi|^2 \Psi = s \frac{\nabla^2 |\Psi|}{|\Psi|} \Psi,\label{RNLS}
\end{equation}
where
\begin{equation}
\nabla^2 = \frac{\partial^2}{{\partial x_1}^2} + \frac{\partial^2}{{\partial x_2}^2} + ... + \frac{\partial^2}{{\partial x_n}^2}.
\end{equation}
With the Madelung representation
$
\Psi = e^{R - i S}
$
there results
\begin{eqnarray}
R_t - [ \nabla^2 S + 2 \nabla R \nabla S] &=& 0, \\
S_t + (1-s) [\nabla^2 R + (\nabla R)^2] - (\nabla S)^2 + \nu e^{2R} &=& 0.
\end{eqnarray}

\subsection{Super-critical case}

For $s > 1$ with
\begin{equation}
\tilde R = R, \hskip0.5cm \tilde S = \frac{S}{\sqrt{s-1}}, \hskip0.5cm \tilde t = t \sqrt{s-1},
\end{equation}
one obtains
\begin{eqnarray}
\tilde R_{\tilde t} - [ \nabla^2 \tilde S + 2 \nabla \tilde R \nabla \tilde S] &=& 0, \\
\tilde S_{\tilde t} - [\nabla^2 \tilde R + (\nabla \tilde R)^2] - (\nabla \tilde S)^2 + \frac{\nu}{s-1} e^{2\tilde R} &=& 0.
\end{eqnarray}
In terms of 
\begin{equation}
e^+ = e^{\tilde R + \tilde S}, \hskip0.5cm e^- = e^{\tilde R - \tilde S}
\end{equation}
it is seen that the system
\begin{eqnarray}
- e^+_{\tilde t} + \nabla^2 e^+ - \frac{\nu}{s-1} (e^+ e^-) e^+ &=&0, \label{RD1}\\
+ e^-_{\tilde t} + \nabla^2 e^- - \frac{\nu}{s-1} (e^+ e^-) e^- &=&0,  \label{RD2}
\end{eqnarray}
results, which admits the 
 conservation law
\begin{equation}
(e^+ e^-)_{\tilde t} + {\bf \nabla} (e^+ {\bf \nabla} e^- - e^- {\bf \nabla} e^+) = 0.
\end{equation}
With
\begin{equation}
|\Psi|^2 = e^+ e^- = e^{2 \tilde R} \equiv \rho
\end{equation}
as fluid density and
\begin{equation}
{\bf v}^- = - \frac{\nabla e^-}{e^-} = - \nabla \ln e^-,
\end{equation}
as the drift velocity, 
the system
\begin{eqnarray}
{{\bf v}^-}_{\tilde t} + \nabla^2 {\bf v}^- &=& \nabla [({\bf v}^-)^2 - \frac{\nu}{s-1} \rho], \label{WBK1}\\
\rho_{\tilde t} - \nabla^2 \rho &=& 2 \nabla (\rho {\bf v}^-), \label{WBK2}
\end{eqnarray}
is obtained.

\subsubsection{Bilinear form}
Real  $g^+$, $g^-$, and $f$ are now introduced according to 
\begin{equation}
e^+ = \frac{g^+}{f}, \hskip0.5cm e^- = \frac{g^-}{f},
\end{equation}
together with Hirota derivatives
\begin{equation}
\frac{\partial}{\partial \tilde t}\, e^{\pm} = \frac{D_{\tilde t} (g^{\pm} \cdot f)}{f^2}.
\end{equation}
For 
\begin{equation}
\nabla = \sum_{i=1}^{n} {\bf e}_i \frac{\partial}{\partial x_i}
\end{equation}
we have
\begin{equation}
\nabla e^\pm = \sum^n_{i=1} {\bf e}_i \frac{\partial}{\partial x_i} \left(\frac{g^\pm}{f}\right) = \sum^n_{i=1} \frac{{\bf e}_i D_{x_i} (g^\pm \cdot f)}{f^2},
\end{equation}
whence, in terms of the vector Hirota derivative operator
\begin{equation}
{\bf D} = \sum^n_{i=1} {\bf e}_i D_{x_i}
\end{equation}
it is seen that
\begin{equation}
\nabla e^\pm = \sum^n_{i=1} {\bf e}_i \frac{\partial}{\partial x_i} \left(\frac{g^\pm}{f}\right) = \frac{{\bf D} (g^\pm \cdot f)}{f^2}.
\end{equation}
In addition,
\begin{equation}
\frac{\partial^2}{{\partial x_i}^2} \, e^{\pm} = \frac{D^2_{x_i} (g^{\pm} \cdot f)}{f^2} - \frac{g^\pm}{f} \frac{D^2_{x_i} (f \cdot f)}{f^2},
\end{equation}
so that, with 
\begin{equation}
\sum^n_{i=1} D^2_{x_i} = D^2_{x_1} + D^2_{x_2} + ... + D^2_{x_n} = {\bf D}^2
\end{equation}
there results
\begin{equation}
\nabla^2 \, e^{\pm} = \frac{ {\bf D}^2 (g^{\pm} \cdot f)}{f^2} - \frac{g^\pm}{f} \frac{{\bf D}^2 (f \cdot f)}{f^2}.
\end{equation}

Bilinear equations corresponding to the system (\ref{RD1}), (\ref{RD2}), are
\begin{eqnarray}
(\mp D_{\tilde t} + {\bf D}^2) (g^{\pm} \cdot f) = 0, \\
{\bf D}^2 (f \cdot f) = - \frac{\nu}{s-1} g^+ g^-,
\end{eqnarray}
where the latter yields
\begin{equation}
\rho = - \frac{2(s-1)}{\nu} \nabla^2 \ln f.
\end{equation}

\subsubsection{Single soliton solution}

With $g^\pm = \epsilon g^\pm_1$, $f = 1 + \epsilon^2 f_2$, and
\begin{equation}
g_1 = e^{\eta^\pm}, \hskip0.5cm \eta^\pm = ({\bf k}^\pm \cdot {\bf x}) + \omega^\pm \tilde t + \eta^\pm_{0},
\end{equation}
there results
\begin{equation}
(\mp \partial_{\tilde t} + \nabla^2) g^{\pm} = 0,
\end{equation}
with solution
\begin{equation}
\omega^\pm = \pm ({\bf k}^\pm)^2.
\end{equation}
Then, for 
$
f = 1 + f_2, $
we obtain
\begin{equation}
\nabla^2 f_2 = - \frac{\nu}{2(s-1)} e^{\eta^+ + \eta^-},
\end{equation}
with solution
\begin{equation}
f_2 = -\frac{\nu}{2(s-1)} \frac{e^{\eta^+ + \eta^-}}{({\bf k}^+ + {\bf k}^-)^2} .
\end{equation}
It may now be established that
\begin{eqnarray}
{\bf D}^2 (f_2,f_2) = c {\bf D}^2 (e^{\eta^+ + \eta^-}, e^{\eta^+ + \eta^-}) &=& 0, \\
(\mp D_{\tilde t} + {\bf D}^2) (e^{\eta^\pm} \cdot e^{\eta^+ + \eta^-}) &=& 0,
\end{eqnarray}
together with the exact solution
\begin{eqnarray}
g^\pm = e^{\eta^\pm}, \hskip0.5cm \eta^\pm = ({\bf k}^\pm \cdot {\bf x} ) \pm ({\bf k}^\pm)^2 \tilde t + \eta^\pm_0,\\
f = 1 - \frac{\nu}{2(s-1)} \frac{e^{\eta^+ + \eta^-}}{({\bf k}^+ + {\bf k}^-)^2},
\end{eqnarray}
wherein  $({\bf k}^\pm \cdot {\bf x} ) = k^\pm_1 x_1 + k^\pm_2 x_2 + ... + k^\pm_n x_n$. Regularity of this solution requires $\nu < 0$,
and we obtain the one dissipaton solution
\begin{equation}
e^\pm ({\bf x}, \tilde t)= e^{\tilde R \pm \tilde S} = \frac{g^\pm}{f} = \frac{e^{\eta^\pm}}{1 + \frac{|\nu|}{2(s-1) ({\bf k}^+ + {\bf k}^-)^2} e^{\eta^+ + \eta^-}}. \label{dissipaton}
\end{equation}
The moving frame 
\begin{equation}
\eta^+ + \eta^- = ({\bf k}^+ + {\bf k}^-) \cdot ( {\bf x} + ({\bf k}^+ - {\bf k}^-) \tilde t - \tilde{\bf x}_0),
\end{equation}
wherein $-({\bf k}^+ + {\bf k}^-) \cdot \tilde {\bf x}_0 \equiv \eta^+_0 + \eta^-_0$, suggests the introduction of the pair of  $n$-dimensional  vectors
\begin{equation}
{\bf k}^+ = {\vec\lambda} + {\vec\mu}, \hskip0.5cm {\bf k}^- = {\vec \lambda} - {\vec \mu},
\end{equation}
whence
\begin{equation}
{\vec\lambda} = \frac{1}{2} ({\bf k}^+ + {\bf k}^-), \hskip0.5cm  {\vec\mu} = \frac{1}{2} ({\bf k}^+ - {\bf k}^-).
\end{equation}
Then, the frame
\begin{equation}
\eta^+ + \eta^- = 2 {\vec\lambda} \cdot ( {\bf x} + 2 {\vec\mu} \,\tilde t - \tilde{\bf x}_0),
\end{equation}
moves with constant velocity vector $\tilde{\bf V} = - 2{\vec\mu}$, so that
\begin{equation}
\eta^+ + \eta^- = 2 {\vec\lambda} \cdot ({\bf x} - \tilde{\bf V} \tilde t - \tilde {\bf x}_0).
\end{equation}
The density $\rho = e^+ e^-$ is given by
\begin{equation}
\rho({\bf x}, \tilde t) = \frac{e^{2 {\vec\lambda} \cdot ({\bf x} - \tilde{\bf V} \tilde t - \tilde {\bf x}_0)}}{\left[1 + \frac{|\nu|}{8(s-1) {\vec \lambda}^2} e^{2 {\vec\lambda} \cdot ({\bf x} - \tilde{\bf V} \tilde t - \tilde {\bf x}_0)}  \right]^2}.
\end{equation}
In terms of 
\begin{equation}
e^{2 {\vec\lambda} \cdot \tilde {\bf x}_1} \equiv \frac{8 (s-1) {\vec\lambda}^2}{|\nu|} 
\end{equation}
and ${\bf x}_0 \equiv \tilde{\bf x}_0 + \tilde{\bf x}_1$,
a soliton traveling wave form results for the density
\begin{equation}
\rho({\bf x}, \tilde t) = \frac{2(s-1)}{|\nu|} \frac{{\vec\lambda}^2}{\cosh^2 {\vec\lambda} \cdot ({\bf x} - \tilde{\bf V} \tilde t - {\bf x}_0) }.
\end{equation}
For drift velocity 
\begin{equation}
{\bf v}^- = - \nabla \ln e^- = - \nabla \ln \frac{g^-}{f}
\end{equation}
we obtain
\begin{equation}
{\bf v}^- = - \nabla \eta^- + \frac{\nabla f}{f}
\end{equation}
and  $\nabla \eta^- = {\bf k}_-$, with
\begin{equation}
\frac{\nabla f}{f} = \frac{|\nu|}{2(s-1)} \frac{({\bf k}^+ + {\bf k}^-)}{({\bf k}^+ + {\bf k}^-)^2} \frac{e^{\eta^+ + \eta^-}}{1 + \frac{|\nu|}{2(s-1) ({\bf k}^+ + {\bf k}^-)^2} e^{\eta^+ + \eta^-}}.
\end{equation}
In terms of vectors ${\vec\lambda}$ and ${\vec\mu}$,
\begin{equation}
\frac{\nabla f}{f} = \frac{|\nu|}{4(s-1)} \frac{\vec\lambda}{{\lambda}^2} \frac{e^{\eta^+ + \eta^-}}{1 + \frac{|\nu|}{8(s-1) {\lambda}^2} e^{\eta^+ + \eta^-}}
\end{equation}
or
\begin{equation} 
\frac{\nabla f}{f} = 2 \vec\lambda \frac{e^{2{\vec\lambda} \cdot ({\bf x} - \tilde{\bf V} \tilde t - {\bf x}_0)}}{1 + e^{2{\vec\lambda} \cdot ({\bf x} - \tilde{\bf V} \tilde t - {\bf x}_0) }}.
\end{equation}
Then for drift velocity the vector shock-soliton form
\begin{equation}
{\bf v}^- ({\bf x}, \tilde t) = {\vec\lambda} \tanh {\vec\lambda} \cdot ({\bf x} - \tilde{\bf V} \tilde t - {\bf x}_0) - \frac{1}{2} \tilde{\bar V}
\end{equation}
results
with asymptotic values at infinity $-\frac{1}{2} \tilde{\bf V} + {\vec\lambda}$ and $-\frac{1}{2} \tilde{\bf V} - {\vec\lambda}$. Vector shock solitonic phenomena and the Hirota method 
are detailed in \cite{P2}.

In summary, the system (\ref{WBK1}), (\ref{WBK2}) admits the exact solution with
\begin{eqnarray}
\rho({\bf x}, \tilde t) = \frac{2(s-1)}{|\nu|} \frac{{\vec\lambda}^2}{\cosh^2 {\vec\lambda} \cdot ({\bf x} - \tilde{\bf V} \tilde t - {\bf x}_0) },\\
{\bf v}^- ({\bf x}, \tilde t) = {\vec\lambda} \tanh {\vec\lambda} \cdot ({\bf x} - \tilde{\bf V} \tilde t - {\bf x}_0) - \frac{1}{2} \tilde{\bar V}.
\end{eqnarray}

 \subsubsection{Envelope soliton of the RNLS equation }

The preceding solution provides an envelope soliton solution of the resonant nonlinear Schr\"odinger (RNLS) equation (\ref{RNLS}) equivalent to (\ref{dissipaton}). 
Thus
\begin{equation}
\tilde S ({\bf x}, \tilde t) = \frac{1}{2} (\eta^+ - \eta^-) = \frac{1}{2} [({\bf k}^+ - {\bf k}^-)\cdot {\bf x} + ({{\bf k}^+}^2 + {{\bf k}^-}^2) \tilde t + \eta^+_0 - \eta^-_0]
\end{equation}
or
\begin{equation}
\tilde S ({\bf x}, \tilde t) = {\vec \mu} \cdot {\bf x} + ({\vec \lambda}^2 + {\vec \mu}^2) \tilde t + \frac{1}{2} [\eta^+_0 - \eta^-_0],
\end{equation}
whence
\begin{equation}
S ({\bf x}, t) = \sqrt{s-1}\, {\vec \mu} \cdot {\bf x} + (s-1) ({\vec \lambda}^2 + {\vec \mu}^2) \, t + \frac{1}{2}\sqrt{s-1}\, [\eta^+_0 - \eta^-_0].
\end{equation}
With $\Psi({\bf x}, t) = \sqrt{\rho} e^{-iS}$, the envelope soliton
\begin{equation}
\Psi({\bf x}, t) = \sqrt{\frac{2(s-1)}{|\nu|}} \frac{|{\vec \lambda}|}{\cosh^2 {\vec\lambda} \cdot ({\bf x} - {\bf V}  t - {\bf x}_0)}
e^{\frac{i}{2} {\bf V} \cdot {\bf x} - i (\frac{1}{4} {\bf V}^2 + (s-1) {\vec \lambda}^2) t - i\phi}
\end{equation}
of the RNLS results in the super-critical case $s > 1$. Here, the notation
${\bf V} = \sqrt{s-1} \tilde{\bf V}$ and  $\phi = (\eta^+_0 - \eta^-_0) \sqrt{s-1}/2$ is adopted.

\subsection{Sub-critical case}

For $s < 1$ with
\begin{equation}
\tilde R = R, \hskip0.5cm \tilde S = \frac{S}{\sqrt{1-s}}, \tilde t = \sqrt{1-s}\, t
\end{equation}
 the system 
\begin{eqnarray}
\tilde R_{\tilde t} - [\nabla^2 \tilde S + 2 \nabla \tilde R \nabla \tilde S] =0,\\
\tilde S_{\tilde t} + [\nabla^2 \tilde R + (\nabla \tilde R)^2] - (\nabla \tilde S)^2 + \frac{\nu}{1-s} e^{2 \tilde R} =0, 
\end{eqnarray}
results.
Then for the wave function $\tilde \Psi ({\bf x}, \tilde t) = e^{\tilde R - i \tilde S}$ an $n+1$ dimensional NLS equation
\begin{equation}
i \tilde \Psi_{\tilde t} + \nabla^2 \tilde \Psi + \frac{\nu}{1-s} |\tilde \Psi|^2 \tilde \Psi = 0
\end{equation}
is retrieved.
Hydrodynamic representation for the latter has been set down in Section 2.3.

\section{A quantum Whitham-Broer-Kaup system in $n+1$ dimensions}

Here the hydrodynamic form of the RNLS equation for drift velocities is considered and a quantum WBK hydrodynamic system connection recorded.
Thus, for the RNLS equation
\begin{equation}
i\Psi_t + \nabla^2 \Psi + \nu |\Psi|^2 \Psi = s \frac{\nabla^2 |\Psi|}{|\Psi|} \Psi, \label{RNLS1}
\end{equation}
the Madelung representation
$
\Psi = e^{R - i S}
$
gives
\begin{eqnarray}
R_t - [ \nabla^2 S + 2 \nabla R \nabla S] &=& 0, \\
S_t + (1-s) [\nabla^2 R + (\nabla R)^2] - (\nabla S)^2 + \nu e^{2R} &=& 0.
\end{eqnarray}
The Madelung representation as a complex Cole-Hopf transformation
\begin{equation}
\frac{\nabla \Psi}{\Psi} = \nabla R - i S = {\bf v}_q + i {\bf v}_c
\end{equation}
introduces the quantum and current velocities
\begin{equation}
{\bf v}_q = \nabla \ln \rho, \hskip0.5cm {\bf v}_c = -2 \nabla S,
\end{equation}
where 
\begin{equation}
\rho = |\Psi|^2 = e^{2R}.
\end{equation}
Then drift velocities are introduced according to 
\begin{eqnarray}
{\bf V}^+ &=& \frac{1}{2} ({\bf v}_q - {\bf v}_c) = \nabla S + \frac{1}{2} \nabla \ln \rho, \\
{\bf V}^- &=& \frac{1}{2} (-{\bf v}_q - {\bf v}_c) = \nabla S - \frac{1}{2} \nabla \ln \rho,
\end{eqnarray}
so that
\begin{eqnarray}
\nabla R &=&  \frac{1}{2} ({\bf V}^+ - {\bf V}^-) =\frac{1}{2} \nabla \ln \rho   ,\\
\nabla S &=& \frac{1}{2} ({\bf V}^+ + {\bf V}^-) = {\bf V}^- + \frac{1}{2} \nabla \ln \rho.
\end{eqnarray}
For drift velocity ${\bf V}^-$ and density $\rho$ the system
\begin{eqnarray}
{\bf V}^-_t + \nabla^2 {\bf V}^- &=& \nabla \left[ ({\bf V}^-)^2 - \nu \rho + (s-2) \frac{\nabla^2 \sqrt{\rho}}{\sqrt{\rho}} \right],\\
\rho_t - \nabla^2 \rho &=& 2 \nabla (\rho {\bf V}^-), \\ rot\, {\bf V}^- &=&0,
\end{eqnarray}
is obtained.
For arbitrary values of the parameter $s \neq 2$ this system represents a quantum WBK hydrodynamic system in $n+1$ dimensions.

In the special case $s =2$, the contribution of the quantum potential vanishes, resulting in the  $n+1$ dimensional WKB system
\begin{eqnarray}
{\bf V}^-_t + \nabla^2 {\bf V}^- &=& \nabla \left[ ({\bf V}^-)^2 - \nu \rho \right],\\
\rho_t - \nabla^2 \rho &=& 2 \nabla (\rho {\bf V}^-), \\ rot\, {\bf V}^- &=&0.
\end{eqnarray}

For the particular case $n=1$ the WBK system 
 \begin{eqnarray}
{V}^-_t + {V}^-_{xx} &=& \left[ ({V}^-)^2 - \nu \rho \right]_x,\\
\rho_t - \rho_{xx} &=& 2  (\rho {V}^-)_x,
\end{eqnarray}
constitutes an
integrable model connected to the classical Boussinesq system
\begin{eqnarray}
{V}^-_t - 2 V^-{V}^-_{x} + p_x =0,\\
p_t + V^-_{xxx} - 2  (p { V}^-)_x = 0.
\end{eqnarray}
Bilinear equations for the latter system are 
\begin{eqnarray}
(D_t + D^2_x) (\tau_1 \cdot \tau_2) =0, \\
D_x(D_t + D^2_x) (\tau_1 \cdot \tau_2) =0,
\end{eqnarray}
where
\begin{equation}
V^- = - (\ln \frac{\tau_1}{\tau_2})_x, \hskip0.5cm \rho = - \frac{2}{\nu} (\ln \tau_2)_{xx}, \hskip0.5cm p = - (\ln (\tau_1 \tau_2))_{xx}.
\end{equation}

\section{Resonant DS system}

Integrable $2+1$-dimensional Whitham-Broer-Kaup systems and resonant NLS connections have been derived in \cite{RogersPashaev}.
A quantum hydrodynamic version is presented 
below commencing with the resonant Davey-Stewartson system
\begin{eqnarray}
i \Psi_t + \nabla^2 \Psi + (\delta -1) \frac{\nabla^2 |\Psi|}{|\Psi|} \Psi + \gamma |\Psi|^2 + \frac{1}{2} \Pi \Psi &=&0,\\
\Pi_{xx} - \Pi_{yy} + 4\gamma (|\Psi|^2)_{xx} &=&0,
\end{eqnarray} 
wherein here  $\delta = 1-s$ and $\gamma = \nu$. 
Under the substitution $\Psi = e^{R - i S}$ the system
\begin{eqnarray}
R_t - [\nabla^2 S  + 2 \nabla S \nabla R] &=&0,\\
S_t + \delta [\nabla^2 R + (\nabla R)^2] - (\nabla S)^2 + \gamma e^{2R} + \frac{1}{2} \Pi &=&0,\\
\Pi_{xx} - \Pi_{yy} + 4\gamma (e^{2R})_{xx} &=&0,
\end{eqnarray}
results.
On introduction of the drift velocities
\begin{equation}
{\bf V}^+ = \nabla S + \frac{1}{2} \nabla \ln \rho,\hskip0.5cm {\bf V}^- = \nabla S - \frac{1}{2} \nabla \ln \rho,
\end{equation}
so that
\begin{equation}
\nabla S = {\bf V}^- + \frac{1}{2} \nabla \ln \rho, \hskip0.5cm R = \frac{1}{2} \ln \rho,
\end{equation}
a WBK type system is obtained, namely
\begin{eqnarray}
{\bf V}^-_t + \nabla^2 {\bf V}^- = \nabla[({\bf V}^-)^2 - \gamma \rho - \frac{1}{2} \Pi - (\delta+1) \frac{\nabla^2 \sqrt{\rho}}{\sqrt{\rho}}], \\
\rho_t - \nabla^2 \rho = 2 \nabla (\rho {\bf V}^-),\\
\Pi_{xx} - \Pi_{yy} + 4\gamma \rho_{xx} =0,
\end{eqnarray}
with $rot \, {\bf V}^- =0$. The system is integrable for any $\delta \neq 1$ and can be connected with the Davey-Stewartson  or resonant Davey-Stewartson 
solitonic equations via appropriate change of variables.
It admits two canonical reductions, namely:

1) when $\delta = -1$($s=2$), the classical hydrodynamic system

\begin{eqnarray}
{\bf V}^-_t + \nabla^2 {\bf V}^- = \nabla[({\bf V}^-)^2 - \gamma \rho - \frac{1}{2} \Pi], \\
\rho_t - \nabla^2 \rho = 2 \nabla (\rho {\bf V}^-),\\
\Pi_{xx} - \Pi_{yy} + 4\gamma \rho_{xx} =0,
\end{eqnarray}
results \cite{RogersPashaev}.

2) when $\delta =1$($s=0$), a novel quantum hydrodynamic system is obtained with

\begin{eqnarray}
{\bf V}^-_t + \nabla^2 {\bf V}^- = \nabla[({\bf V}^-)^2 - \gamma \rho - \frac{1}{2} \Pi - 2 \frac{\nabla^2 \sqrt{\rho}}{\sqrt{\rho}}], \\
\rho_t - \nabla^2 \rho = 2 \nabla (\rho {\bf V}^-),\\
\Pi_{xx} - \Pi_{yy} + 4\gamma \rho_{xx} =0.
\end{eqnarray}

In the classical case, the bilinear representation and admitted solitons have been detailed by Rogers and Pashaev in \cite{RogersPashaev}, while the dressing method 
has subsequently been applied by Nabelik and Zakharov in \cite{NabelekZakharov}. In the quantum case, a spectrum of solutions may be generated via the Davey-Stewartson-I 
equation, which notably, admits exponentially localized solutions commonly termed dromions.   The solutions can simulate quantum effects as inelastic scattering, 
fusion and fission, creation and annihilation.

\section{The $n+1$-dimensional super-critical resonant NLS equation: Painlev\'e XXXIV symmetry reduction}

Here, a Painlev\'e XXXIV symmetry reduction of the $n+1$-dimensional resonant NLS equation 

\begin{equation}
i \Psi_t + \nabla^2 \Psi + \nu |\Psi|^2 \Psi - s \frac{\nabla^2 |\Psi|}{|\Psi|}\Psi =0 \label{RNLS1}
\end{equation}
is derived subject to the super critical condition $s > 1$. The induced such reduction for an equivalent multi-dimensional extension of the Whitham-Broer-Kaup system is subsequently 
thereby obtained.

Thus, the ansatz with
\begin{equation}
\Psi = [\phi(\xi) + i \psi(\xi)] e^{i\chi} \label{ansatz}
\end{equation}
\begin{equation}
\xi = \alpha t + \beta t^2 + x_1 + x_2 + ... + x_n,\label{xi}
\end{equation}

\begin{equation}
\chi = \gamma t^3 + \delta t^2 + \epsilon t (x_1 + x_2 + ... + x_n) + \zeta t + \lambda_1 x_1 +\lambda_2 x_2 + ... \lambda_n x_n,\label{chi}
\end{equation}
is adopted wherein $\alpha, \beta, \gamma, \delta, \epsilon, \zeta$ and $\lambda_i$, $i=1,2,...,n$ are real constants.
Insertion into the resonant NLS equation (\ref{RNLS1}) on separation of real and imaginary parts respectively yields
\begin{eqnarray}
n \phi'' - \psi' [\alpha + 2 \beta t + 2 n \epsilon t + 2 (\lambda_1 + \lambda_2 + ... + \lambda_n)] \nonumber \\
- \phi [3 \gamma t^2 + 2 \delta t + \epsilon (x_1 + x_2 + ... + x_n) + \zeta] \nonumber \\
- \phi [(\epsilon t + \lambda_1)^2 + (\epsilon t + \lambda_2)^2 + ... + (\epsilon t + \lambda_n)^2] \nonumber \\
- \nu (\phi^2 + \psi^2) \phi - n s \frac{|\psi|_{\xi\xi}}{|\psi|} \phi =0,\label{Real}
\end{eqnarray}

\begin{eqnarray}
n \psi'' + \phi' [\alpha + 2 \beta t + 2 n \epsilon t + 2 (\lambda_1 + \lambda_2 + ... + \lambda_n)] \nonumber \\
- \psi [3 \gamma t^2 + 2 \delta t + \epsilon (x_1 + x_2 + ... + x_n) + \zeta] \nonumber \\
- \psi [(\epsilon t + \lambda_1)^2 + (\epsilon t + \lambda_2)^2 + ... + (\epsilon t + \lambda_n)^2] \nonumber \\
+ \nu (\phi^2 + \psi^2) \psi - n s \frac{|\psi|_{\xi\xi}}{|\psi|} \psi =0,\label{Imaginary}
\end{eqnarray}
Accordingly,
\begin{equation}
n \phi'' \psi - n \psi'' \phi - (\phi \phi' + \psi \psi')[\alpha + 2 \beta t + 2 n \epsilon t + 2 (\lambda_1 + \lambda_2 + ... + \lambda_n)] = 0
\end{equation}
whence, it is required that $\beta = - n \epsilon$ and, on integration, there results a key integral of motion, namely
\begin{equation}
n ( \phi' \psi - \phi \psi') - \frac{1}{2} (\phi^2 + \psi^2)
[\alpha + 2(\lambda_1 + \lambda_2 + ... + \lambda_n)] = I,\,\,\,I \in R \label{keyintegral}
\end{equation}

On use of the $\xi$-relation in (\ref{xi}) in the real and imaginary decompositions independence therein of the terms involving $t$ and $t^2$ requires that
\begin{eqnarray}
2[\delta + \epsilon (\lambda_1 + \lambda_2 + ,,,+ \lambda_n)] &=& \alpha \epsilon, \\
3 \gamma + n \epsilon^2 &=& \beta \epsilon
\end{eqnarray}

whence, reduction of (\ref{Real}), (\ref{Imaginary}) is obtained to
\begin{eqnarray}
n \phi'' - \psi'[\alpha + 2 (\lambda_1 + \lambda_2 + ...+ \lambda_n)] - \phi [\zeta + \epsilon \xi + \lambda^2_1 + \lambda^2_2 + ... \lambda^2_n] \nonumber \\
- \nu (\phi^2 + \psi^2) \phi - n s \frac{|\psi|_{\xi\xi}}{|\psi|} \phi = 0,
\end{eqnarray}
\begin{eqnarray}
n \psi'' + \phi'[\alpha + 2 (\lambda_1 + \lambda_2 + ...+ \lambda_n)] - \psi [\zeta + \epsilon \xi + \lambda^2_1 + \lambda^2_2 + ... \lambda^2_n] \nonumber \\
+ \nu (\phi^2 + \psi^2) \psi - n s \frac{|\psi|_{\xi\xi}}{|\psi|} \psi = 0.
\end{eqnarray}
The latter system produces the relation
\begin{eqnarray}
n(\phi \phi'' + \psi \psi'') + (\phi' \psi - \psi' \phi) [\alpha + 2 (\lambda_1 + \lambda_2 + ...+ \lambda_n)] \nonumber \\
- (\phi^2 + \psi^2)[\zeta + \epsilon \xi + \lambda^2_1 + \lambda^2_2 + ... + \lambda^2_n]\nonumber \\
+ \nu (\phi^2 + \psi^2)^2 - n s \frac{|\psi|_{\xi\xi}}{|\psi|} (\phi^2 + \psi^2) = 0 \label{relation0}
\end{eqnarray}
wherein
\begin{equation}
|\psi|_{\xi\xi} = (\phi \phi'' + \psi \psi'' + {\phi'}^2 + {\psi'}^2) |\psi|^{-1} - (\phi \phi' + \psi \psi')^2 |\psi|^{-3}.
\end{equation} 
This relation together with the identity
\begin{equation}
(\phi^2 + \psi^2)(\phi'^2 + \psi'^2) - (\phi' \psi - \psi' \phi)^2 \equiv (\phi \phi' + \psi \psi')^2
\end{equation}
and the integral of motion $I$ show that
\begin{equation}
|\psi|_{\xi\xi} |\psi|^3 = (\phi'' \phi + \psi'' \psi) |\psi|^2 + \frac{1}{n^2}[I + \frac{1}{2} |\psi|^2 
(\alpha + 2 (\lambda_1 + \lambda_2 +...+\lambda_n))]^2.
\end{equation}
Elimination of $\phi'' \phi + \psi'' \psi$ in the latter on use of (\ref{relation0}) now yields
\begin{eqnarray}
|\psi|_{\xi\xi} |\psi|^3 = \frac{1}{n} [-(\phi' \psi - \psi' \phi)[\alpha + 2 (\lambda_1 + \lambda_2 +...+\lambda_n)] \nonumber \\
+ |\psi|^2 (\zeta + \epsilon \xi + \lambda^2_1 + \lambda^2_2 + ... \lambda^2_n) - \nu |\psi|^4 + n s |\psi|_{\xi\xi} |\psi| \,] |\psi|^2
\nonumber \\
\frac{1}{n^2}[I + \frac{1}{2} |\psi|^2 
(\alpha + 2 (\lambda_1 + \lambda_2 +...+\lambda_n))]^2
\end{eqnarray}
whence reduction is obtained to
\begin{eqnarray}
(1-s)|\psi|_{\xi\xi} |\psi|^3 = -\frac{1}{4 n^2} [\alpha + 2 (\lambda_1 + \lambda_2 +...+\lambda_n)]^2 |\psi|^4 \nonumber \\
+ \frac{\epsilon}{n} \xi |\psi|^4 +\frac{1}{n} (\zeta +  \lambda^2_1 + \lambda^2_2 + ... \lambda^2_n) |\psi|^4
\nonumber \\
- \frac{\nu}{n} |\psi|^6 + \frac{1}{n^2} I^2.
 \end{eqnarray}
Accordingly, 
\begin{equation}
|\psi|_{\xi\xi} + [\eta - \frac{\epsilon \xi}{n (1-s)}] |\psi| + \frac{\nu}{n (1-s)}|\psi|^3 = \frac{\cal E^*}{|\psi|^3}
\end{equation}
wherein
\begin{equation}
\eta = \frac{1}{(1-s)} [\frac{1}{4 n^2} (\alpha + 2 (\lambda_1 + \lambda_2 +...+\lambda_n))^2 - \frac{1}{n} (\zeta +  \lambda^2_1 + \lambda^2_2 + ... \lambda^2_n)  ]
\end{equation}
together with
\begin{equation}
{\cal E^*} = \frac{1}{1-s} \left(\frac{I}{n}\right)^2 < 0, \hskip0.5cm (s > 1)
\end{equation}
Under the translation $\xi^* = \xi + n \eta (s-1)/\epsilon$ with $\epsilon \neq 0$ a hybrid Ermakov-Painlev\'e II equation \cite{Rogers2}, namely
\begin{equation}
|\psi|_{\xi^*\xi^*} + \epsilon^* \xi^* |\psi| + \nu^* |\psi|^3 = \frac{\cal E^*}{|\psi|^3}
\end{equation}
wherein $\epsilon^* = \epsilon/n (s-1)$ and $\nu^* = \nu/n(s-1)$. The substitution $|\psi| = \Sigma^{1/2}$, $\Sigma > 0$ produces, on 
appropriate scalings, a Painlev\'e XXXIV canonical reduction
\begin{equation}
\Sigma_{\xi^*\xi^*} - \frac{1}{2} \frac{\Sigma^2_{\xi^*}}{\Sigma} + 2 \Sigma^2 - \xi \Sigma = - \frac{(\alpha^* + \frac{1}{2})^2}{2\Sigma}, 
\,\,\alpha^*\in R \label{Painleve}
\end{equation}
Iterative action of a B\"acklund transformations admitted as in \cite{RogersSchief3} generates sequences of exact solutions of (\ref{Painleve}) and, accordingly, in turn of the $n+1$-dimensional 
supercritical NLS equation (\ref{RNLS1}).

\section{An induced $n+1$-dimensional Whitham-Broer-Kaup system: Painlev\'e reduction}

On introduction of the Madelung decomposition $\Psi = e^{R - iS}$ into $n+1$-dimensional resonant NLS equation (\ref{RNLS1}) there results
\begin{eqnarray}
R_t - \sum^n_{i=1} [S_{x_i x_i} + 2 R_{x_i} S_{x_i}] = 0, \label{RS1}\\
S_t + \sum^n_{i=1} [(1-s) (R_{x_i x_i} + R^2_{x_i}) - S^2_{x_i}] + \nu |\Psi|^2 = 0.\label{RS2}
\end{eqnarray}
Thus, with the super-critical resonance condition $s> 1$ and 
\begin{eqnarray}
R^* = R, \hskip0.5cm S^* = (s-1)^{-1/2} S, \hskip0.5cm t^* = (s-1)^{1/2} t,
\end{eqnarray}
the system (\ref{RS1}), (\ref{RS2}) becomes
\begin{eqnarray}
R^*_{t^*} - \sum^n_{i=1} [S^*_{x_i x_i} + 2 R^*_{x_i} S^*_{x_i}] = 0, \label{RS21}\\
S^*_{t^*} + \sum^n_{i=1} [ R^*_{x_i x_i} + {R^*_{x_i}}^2 - {S^*_{x_i}}^2] + \frac{\nu}{s-1} |\Psi|^2 = 0.\label{RS22}
\end{eqnarray}
On introduction of 
\begin{equation}
e^+ = e^{R^* + S^*}, \hskip1cm e^- = e^{R^* - S^*},
\end{equation}
the preceding shows that
\begin{equation}
- e^+_{t^*} +\nabla^2 e^+ - \frac{\nu}{s-1} [e^+ e^-] e^+ =0, 
\end{equation}
together with
\begin{equation}
 e^-_{t^*} +\nabla^2 e^- - \frac{\nu}{s-1} [e^+ e^-] e^- =0.
\end{equation}
Thus, 
\begin{equation}
e^-_{t^*} e^+ + e^+_{t^*} e^- = - (\nabla^2 e^-) e^+ + (\nabla^2 e^+) e^-
\end{equation}
so that
\begin{equation}
(e^+ e^-)_{t^*} = \sum^n_{i=1} [(e^+ e^-)_{x_i} + 2 (e^+ e^-) \left(- \frac{e^-_{x_i}}{e^-} )\right)]_{x_i}
\end{equation}
Hence, on introduction of 
\begin{equation}
h = e^+ e^- = |\Psi|^2, \hskip0.5cm \theta_i = - \frac{e^-_{x_i}}{e^-}; i = 1,2,...,n
\end{equation}
there results
\begin{equation}
h_{t^*} - \nabla^2 h -2 \sum^n_{i=1} (h \theta_i)_{x_i} = 0
\end{equation}
while
\begin{eqnarray}
\theta_{i, t^*} = - \left(\frac{e^-_{x_i}}{e^-}\right)_{t^*} = - \left(\frac{e^-_{t^*}}{e^-}\right)_{x_i} \nonumber \\
= \left[ \frac{\nabla^2 e^-}{e^-} - \frac{\nu}{s-1} |\Psi|^2\right]_{x_i} = \sum^{n}_{i=1} [- (\theta_i)_{x_i} + \theta^2_i]_{x_i} 
- \frac{\nu}{s-1}\left[ |\Psi|^2\right]_{x_i}
\end{eqnarray}
$i= 1,2,...,n$.

\hskip5cm  Summary

Thus, a coupled multi-dimensional system equivalent to the $n+1$-dimensional super-critical NLS system (\ref{RNLS1}) has been derived, namely
\begin{eqnarray}
h_{t^*} - \nabla^2 h -2 \sum^n_{i=1} (h \theta_i)_{x_i} &=& 0, \label{H1}\\
\theta_{i, t^*} + \sum^{n}_{i=1} [ (\theta_i)_{x_i} - \theta^2_i]_{x_i} 
- \frac{\nu}{s-1} h_{x_i} &=&0,\,\,\,\, i=1,2,...,n\label{H2}
\end{eqnarray}
wherein 
\begin{equation}
h = |\Psi|^2, \hskip0.5cm \theta_i = - \frac{e^-_{x_i}}{e^-}
\end{equation}
and
\begin{equation}
\theta_{i, x_j} = \theta_{j, x_i}
\end{equation}

\section{Specializations of the System (\ref{H1})-(\ref{H2})}

\subsection{The Classical Whitham-Broer-Kaup System}
The $1+1$-reduction of the nonlinear coupled system (\ref{H1}), (\ref{H2}) produces the canonical Whitham-Broer-Kaup hydrodynamic system
\begin{eqnarray}
h^*_{t^*} -  h^*_{xx} +2 (h^* \theta^*)_{x} = 0, \label{H11}\\
\theta^*_{t^*} + \theta^*_{x x} + 2 \theta^*\theta^*_x - h^*_x =0 \label{H12}
\end{eqnarray}
wherein
\begin{equation}
h^* = \frac{\nu}{s-1} h = \frac{\nu}{s-1} |\Psi|^2, \hskip0.5cm \theta^* = -\theta_1, \hskip0.5cm t^* = (s-1)^{1/2} t
\end{equation}
\subsubsection{$s>1$}
The super-critical condition $s > 1$ corresponds to the condition that the $1+1$-dimensional resonant NLS equation admits solitonic fusion and fission phenomena \cite{PashaevLeeRogers}. Accordingly, the admittance of such characteristics is inherited  by the equivalent  Whitham-Broer-Kaup system (\ref{H11}),(\ref{H12}). Indeed such resonance behaviour admitted by that integrable nonlinear system has been indicated via a 
trilinear representation in \cite{Satsuma}.

\subsection{A Novel $2+1$-Dimensional Symmetric  Whitham-Broer-Kaup System}
The $2+1$-dimensional reduction of the system (\ref{H1}),(\ref{H2}) connects the super-critical resonant nonlinear Schr\"dinger equation
\begin{equation}
i \Psi_t + \Psi_{xx} + \Psi_{yy} + \nu |\Psi|^2 \Psi - s \left(\frac{|\Psi|_{xx}}{|\Psi|} + \frac{|\Psi|_{yy}}{|\Psi|} \right)\Psi =0
\end{equation}
to a novel associated  Whitham-Broer-Kaup type system, namely
\begin{eqnarray}
h^*_{t^*} - h^*_{xx} - h^*_{yy} - 2 [(h^* \theta_1)_x +(h^* \theta_2)_y ] &=&0, \label{21WBK1}\\
\theta_{1, t^*} + [\theta_{1,x} + \theta_{2,y} - \theta^2_1 - \theta^2_2]_x  + h^*_x &=&0, \label{21WBK2}\\
\theta_{2, t^*} + [\theta_{1,x} + \theta_{2,y} - \theta^2_1 - \theta^2_2]_y  + h^*_y &=&0, \label{21WBK3}\\
\theta_{1,y} &=& \theta_{2,x} \label{21WBK4}
\end{eqnarray}
wherein $h^* = (\nu/(s-1)) h$. This system is seen to admit the distinctive property that it is invariant under the interchanges
\begin{equation}
x \leftrightarrow y, \hskip0.5cm \theta_1 \leftrightarrow \theta_2.
\end{equation}

Here, on introduction of $\phi^+$ according to 
\begin{equation}
\theta_1 = \phi^+_x, \hskip0.5cm \theta_x = \phi^+_y
\end{equation}
the $2+1$-dimensional symmetric WBK system (\ref{21WBK1})-(\ref{21WBK4}) yields
\begin{equation}
\phi^+_{t^*} + \phi^+_{xx} + \phi^+_{yy} - {\phi^+_x}^2 - {\phi^+_y}^2 + h^* = T^*(t^*).
 \end{equation}
Thus, with 
\begin{equation}
\phi^+ = \phi^* + \int T^*(t^*) dt^*,
\end{equation}
reduction of (\ref{21WBK1})-(\ref{21WBK4}) is made to the nonlinear system
\begin{eqnarray}
h^*_{t^*} - h^*_{xx} - h^*_{yy} - 2[(h^* \phi^*_x)_x + (h^* \phi^*_y)_y] &=&0, \\
\phi^*_{t^*} + \phi^*_{xx} + \phi^*_{yy}   - {\phi^*_x}^2 - {\phi^*_y}^2 + h^* &=&0.
\end{eqnarray}
The latter constitutes a novel symmetric extension to $2+1$-dimension of the classical Whitham-Broer-Kaup system. It is recalled that the constituent members of the $2+1$-dimensional LKR solitonic system as introduced in \cite{KR1}, \cite{KR2}, likewise are characterized by containing spatial variables $x, y$ with equal standing as in the canonical NVN and Davey-Stewartson systems and the novel symmetric sine-Gordon system of \cite{KR2}.

\subsection{Induced Painlev\'e XXXIV reduction: The n+1-dimensional Whitham-Broer-Kaup system}

The Painleve' XXXIV symmetry reduction (\ref{ansatz}) of the $n+1$-dimensional resonant NLS equation (\ref{RNLS1}) with the super-critical condition $s > 1$ induces an associated reduction for the equivalent Whitham-Broer-Kaup system via the relations
\begin{equation}
h = |\Psi|^2 = \Sigma, \hskip0.5cm \theta_i = - \frac{e^-_{x_i}}{e^-}
\end{equation}
wherein 
\begin{equation}
e^- = e^{R - (s-1)^{-1/2} S}, \hskip0.5cm s>1
\end{equation}
corresponding to the Madelung representation $\Psi = e^{R-iS}$. Thus, similarity ansatz determined by the relations (\ref{ansatz})-(\ref{chi}) requires
\begin{eqnarray}
e^R \cos S = \phi \cos \chi - \psi \sin \chi, \hskip0.5cm e^R \sin S = - [\phi \sin \chi + \psi \cos \chi]
\end{eqnarray}
whence 
\begin{equation}
S = - [\tan^{-1} \frac{\psi}{\phi} + \chi].
\end{equation}

Here, the integral of motion (\ref{keyintegral}) yields
\begin{equation}
- n \frac{d}{d \xi}[\tan^{-1} (\frac{\psi}{\phi})] - \frac{1}{2} [\alpha + 2 (\lambda_1 + \lambda_2 +...+ \lambda_n)] = \frac{I}{|\Psi|^2}
\end{equation}
whence, on integration
\begin{equation}
 n \tan^{-1} (\frac{\psi}{\phi}) + \frac{1}{2} [\alpha + 2 (\lambda_1 + \lambda_2 +...+ \lambda_n)] = - I\int \frac{1}{\Sigma} d\xi + C,\label{relation}
\end{equation}
$(C \in R)$.
Accordingly, 
\begin{equation}
\phi(\xi) = \pm \Sigma^{1/2} \frac{\Delta}{(1 + \Delta^2)^{1/2}}, \hskip0.5cm 
\psi(\xi) = \pm \Sigma^{1/2} \frac{1}{(1 + \Delta^2)^{1/2}}
\end{equation}
with $\Delta = \phi/\psi$ determined by the relation (\ref{relation}). In addition,
\begin{equation}
e^- = |\Psi| \exp \left((s-1)^{-1/2} [\tan^{-1} \frac{\psi}{\phi} + \chi]\right)
\end{equation}
whence the relations
\begin{eqnarray}
\theta_i = - (\ln e^-)_{x_i} = -[\ln |\Psi| + (s-1)^{-1/2} (\tan^{-1} \frac{\psi}{\phi} + \chi)]_{x_i} \\
= - \frac{(\Sigma^{1/2})_\xi}{\Sigma^{1/2}} - (s-1)^{-1/2} (\frac{1}{2n} (\lambda_1 + \lambda_2 + ...+ \lambda_n) + \frac{I}{n \Sigma}
+ \epsilon k + \lambda_i)
\end{eqnarray}
$i =1, 2, ..., n$, completes the symmetry reduction of the $n+1$-dimensional Whitham-Broer-Kaup system in terms of the classical Painleve' XXXIV in $\Sigma$.

\subsection{Integrable spatial modulation of the canonical 1+1-dimensional resonant NLS equation}

The solitonic $1+1$-dimensional resonant NLS equation
\begin{equation}
i \Psi_t + \Psi_{xx} + \nu |\Psi|^2 \Psi - s \frac{|\Psi|_{xx}}{|\Psi|} \Psi = 0 \label{RNLS1dim}
\end{equation}
has important physical applications, notably in cold plasma physics \cite{LeePashaevRogersSchief}. Its temporal modulation has been extensively investigated in the literature.
In \cite{RogersSM}, spacially modulated versions of coupled solitonic systems of sine-Gordon Demoulin and Manakov-type have been derived as generated by 
the application of a class of involutory transformation to their S-integrable unmodulated counterparts. In the case of the modulated Manakov system, novel bright 
solitonic phenomena was thereby isolated. Here, such involutory transformations are applied to (\ref{RNLS1dim}) to generate associated modulated versions which
inherit its S-integrable properties.

Thus, the class of transformations
\begin{eqnarray}
\Psi^* = \Psi/\rho(x), \hskip0.5cm dx^* = \rho^{-2}(x) dx, \hskip0.5cm dt^* = dt \,\,:\, I^*
\end{eqnarray}
is introduced. Herein, $I^*$ is seen to admit the important involutary property $I^* = I^{* -1}$ since 
\begin{eqnarray}
\Psi^{**} &=& \Psi^*/\rho^* = \Psi, \hskip0.5cm dx^{**} = \rho^{*-2}(x) dx^* = dx, \\ dt^{**} &=& dt^* = dt, 
\hskip0.5cm
\rho^{**} = \rho.
\end{eqnarray}
Under $I^*$ accordingly, a reciprocal class of spatially modulated $1+1$-dimensional resonant NLS equations results 
related to its canonical solitonic unmodulated counterpart (\ref{RNLS1dim}) in an involutary manner, namely
\begin{eqnarray}
i \left(\frac{\Psi^*}{\rho^*}\right)_{t^*} + {\rho^*}^2 \frac{\partial}{\partial x^*} [{\rho^*}^2 \frac{\partial}{\partial x^*}] \left(\frac{\Psi^*}{\rho^*}\right)
+ \nu \rho^{* -3} |\Psi^*|^2 \Psi^* \nonumber \\
- s \left[ \frac{{\rho^*}^2 \frac{\partial}{\partial x^*} ({\rho^*}^2 \frac{\partial}{\partial x^*}) (|\Psi^*|/\rho^*)}{(|\Psi^*|/\rho^*)}\right] \left(\frac{\Psi^*}{\rho^*}\right) =0
\end{eqnarray}
Ermakov Modulation. 

The classical single component Ermakov equation \cite{Ermakov}
\begin{equation}
\rho^*_{x^* x^*} + \omega^* (x^*) \rho^* = \frac{\cal E}{{\rho^*}^3}, \hskip0.5cm {\cal E} \in R
\end{equation}
admits an important nonlinear superposition principle, namely
\begin{equation}
\rho^* = \sqrt{c_1 \Omega^2_1 + 2 c_2 \Omega_1 \Omega_2 + c_3 \Omega^2_2} \label{nonlinsuperposition}
\end{equation}
wherein $\Omega_1$, $\Omega_2$ constitute a pair of linearly independent solutions of the auxiliary base equation
\begin{equation}
\Omega_{x^* x^*} + \omega^* (x^*) \Omega =0.
\end{equation}
Herein, the $c_i$, $i=1,2,3$ are real constants such that $c_1 c_3 - c^2_2 = {\cal E}/W^2$,
where $W =  \Omega_1 {\Omega_2}_{x^*} - {\Omega_1}_{x^*} \Omega_2$ is the constant Wronskian of $\Omega_1$, $\Omega_2$.

The nonlinear superposition principle (\ref{nonlinsuperposition}) may be retrieved via a Lie group procedure as in 
\cite{RogersR} wherein application was made to solve a class of initial value problems descriptive of moving 
shoreline  evolution in rotating shallow water hydrodynamics. Extended Ermakov-type equations and their integrable 
discretisations which admit nonlinear superposition principles were subsequently derived via Lie group methods in \cite{RogersSW}.
The classical Ermakov superposition principle (\ref{nonlinsuperposition}) has application to the exact solution of a range of initial value 
problems descriptive of the large amplitude radial oscillation of thin-shelled tubes of hyperelastic Mooney-Rivlin material subject to various boundary loadings \cite{RogersAn}

In \cite{RogersSM}, both Ermakov and Ermakov-Painlev\'e integrable modulated versions of coupled solitonic systems of sine-Gordon, Demoulin  and Manakov type have been derived by application of classes of involutary transformations to their unmodulated canonical counterparts. It is 
remarked that the type of reciprocal transformation applied had its genesis in an autonomisation procedure introduced in \cite{AthRogersRO}
for the two-component Ermakov-Ray-Reid extension of the classical Ermakov equation. This generalization has proved to have extensive physical application (qv \cite{RogersSchief2}), notably in nonlinear optics  \cite{RogersMalomed1}, \cite{RogersMalomed2}. 
Nonlinear Schr\"odinger equations with spatial modulation associated with integrable Hamiltonian systems of Ermakov-Ray-Reid type were detailed in \cite{RogersSV}.


\begin{thebibliography}{6}
%
\bibitem{W} Whitham, G.B., Variational methods and applications to water waves, Proc. Roy. Soc. London A299, (1967) 6-25

\bibitem{Broer} Broer, L.J.F., Approximate equations for long water waves, Appl. Sci. Res. 31, (1975) 377-385

\bibitem{Kaup} Kaup, D. J., A higher-order water wave equation and the method for solving it, Prog. Theor. Phys. 54, (1975) 396-408

\bibitem{RogersPashaev} Rogers, C. and Pashaev, O. K., On a 2+1 dimensional Whitham-Broer-Kaup system: a resonant NLS connection, Stud. Appl. Math. 127, (2011) 141-152

\bibitem{RogersYipChow} Rogers, C., Yip, L. P. and  Chow, K.W., A resonant Davey-Stewartson capillarity model system, Int. J. Nonlinear Sci. Num. Simulations, 10 (2009) 397-403 

\bibitem{NabelekZakharov} Nabelek, P. V. and  Zakharov, V. E., Solutions to the Kaup-Broer system and its integrable generalization via the dressing method, Physica D: Nonlinear Phenomena 
409, (2020) 132478

\bibitem{ModPhysLettA} Pashaev, O.K. and Lee, J.-H., Resonance solitons as black holes in Madelung fluid,   Mod. Phys. Lett. A 17, (2002) 1601-1619

\bibitem{PashaevLeeRogers} Pashaev, O. K., Lee, J.-H. and  Rogers, C., Soliton resonances in a generalized nonlinear Schr\"odinger equations, J. Phys. A: Math. Theor., 41, (2006) 45201 (9pp)

\bibitem{MalomedStenflo} Malomed, B. A. and  Stenflo, L., Modulation instabilities and soliton solutions of a generalized nonlinear Schr\"odinger equation, J. Phys. A: Math. Gen. 24,  (1991) 
L1149-1153

\bibitem{RogersSchief1} Rogers, C. and Schief, W. K., The resonant nonlinear Schr\"odinger equation via an integrable capillary model, Il Nuovo Cimento B114, (1999) 1409-1412 

\bibitem{Rogers1} Rogers, C.,  Reciprocal gausson phenomena in a Korteweg capillary system, Meccanica, 54, (2019) 1515-1529

\bibitem{LeePashaevRogersSchief} Lee, J.-H., Pashaev, O.K.,  Rogers, C. and  Schief, W. K.,  The resonant nonlinear Schr\"odinger equation in cold plasma physics. Application of B\"acklund-Darboux 
transformations andsuperposition principle, J. Plasma Physics, 73, (2007) 257-272

\bibitem{Kudryashov} Kudryashev N.A., Optical solitons of the resonant nonlinear Schr\"odinger equation with arbitrary index, Optik, 235, (2021) 160626 

\bibitem{P1} Pashaev O.K, Envelope soliton resonances and Broer-Kaup-type non-Madelung fluids, Theor. Math. Phys., 172 (2), 2012, 1147-1159

\bibitem{Lee} Lee J.-H. and Pashaev O.K, Resonant dispersive Benney and Broer-Kaup systems in 2+1 dimensions, Theor. Math. Phys., 172 (2), 2012, 1147-1159

\bibitem{Giannini} Giannini, J. A. and Joseph R.I., The role of the second Painlev\'e transcendent in nonlinear optics, Phys. Lett. A 141, (1989) 417-419

\bibitem{Rogers2} Rogers, C.  A novel Ermakov-Painlev\'e II system. N+1 dimensional coupled NLS and elastodynamic reductions, Stud. Appl. Math., 133, (2014) 214-231

\bibitem{Ermakov} Ermakov, V. P., Second-order differential equations: conditions of complete integrability, Univ. Izv. Kiev 20, (1880) 1-25 

\bibitem{RogersMalomed1} Rogers, C., Malomed, B., Chow K. W. and An, H., Ermakov-Ray-Reid systems in nonlinear optics, J. Phys. A: Math. Theor. 43 (2010) 455214 (15pp)

\bibitem{RogersMalomed2} Rogers, C.,  Malomed, B. and  An, H., Ermakov-Ray-Reid reductions of variational approximations in nonlinear optics, Stud. Appl. Math. 129 (2012) 389-413 

\bibitem{RogersSchief2} Rogers, C. and  Schief, W. K.,  Ermakov-type systems in nonlinear physics and continuum mechanics, in Nonlinear Systems  and Their Remarkable Mathematical Structures, 
Editor Norbert Euler, CRC Press, (2018) 541-576  


\bibitem{RogersSchief3} Rogers, C. and  Schief, W. K., On Ermakov-Painleve' systems. Integrable reduction, Meccanica 51, (2016) 2967-2974


\bibitem{RogersClarkson1} Rogers, C. and  Clarkson P. A.,  Ermakov-Painlev\'e II symmetry reduction of a Korteweg capillary system, Symmetry, Integrability and Geometry: Methods and Applications, 
13, (2017) 018

\bibitem{RogersClarkson2} Rogers, C. and  Clarkson P. A., Ermakov-Painlev\'e reduction in cold plasma physics. Application of a B\"acklund transformations, J. Nonlinear Mathematical Physics, 25, 
(2018) 242-261

\bibitem{ARogers} Amster P. and Rogers, C., On a Ermakov-Painlev\'e II reduction in three-ion electrodiffusion - A Dirichlet boundary value problem, Discrete and Continuous Dynamical Systems, 35, (2015) 3277-3292 

\bibitem{Zakharov} Zakharov, V.E., On the Benney equations, Physica D3, (1981), 193-202

\bibitem{Maruno} Maruno, K. and Ohta, Y., Localized solitons of a (2+1)-dimensional nonlocal nonlinear Schr\"odinger equation,Phys. Lett. A 372, (2008), 4446-4450

\bibitem{LeePashaev} Lee, J.-H. and  Pashaev, O.K., Chiral resonant solitons in Chern-Simons theory and Broer-Kaup type new hydrodynamic system,  Chaos, Solitons and Fractals, 45 (2012) 1041-1047.

\bibitem{P2} Pashaev, O.K and Tanoglu, G., Vector shock soliton and the Hirota bilinear method, Chaos, Solitons and Fractals, 26, (2005), 95-105
 
 \bibitem{Satsuma} Satsuma J., Kajiwara K. and Matsukidaira J., Solutions of the Broer-Kaup system through its trilinear form, J. Phys. Soc. Japan, 61, (1992) 3096-3102

\bibitem{KR1} Konopelchenko B and Rogers C.,  On 2+1-dimensional nonlinear systems of Loewner-type , Phys. Lett. A 158, (1991) 391-397.

\bibitem{KR2} Konopelchenko B and Rogers C.,  On generalized Loewner systems: novel integrable equations in 2+1-dimensions, J. Math. Phys. 34 (?), (1993) 214-242.

\bibitem{RogersSM} Rogers, C., Schief W.K. and Malomed B., On modulated coupled systems. Canonical reduction via reciprocal transformations, Comm. Nonlinear Sci. and Numerical Simulations, 83, 
(2020) 105091.


\bibitem{RogersR} Rogers, C. and Ramgulam U., A nonlinear superposition principle and Lie group invariance: application to rotating shallow water theory, Int. J. Nonlinear Mech. , 24, (1989) 229-236

\bibitem{RogersSW} Rogers, C., Schief W.K. and Winternitz P., Lie theoretical generalization and discretization of the Pinney equation, J. Math. Anal. Appl., 216 (1997) 246-264


\bibitem{RogersAn} Rogers, C. and Ames W.F., Nonlinear Boundary Value Problems in Science and Engineering, Academic Press, New York, (1989)

\bibitem{AthRogersRO} Athorne C., Rogers, C., Ramgulam U. and Osbaldestin A., On linearization of the Ermakov system, Phys. Lett. A 143 
(1990) 207-212


\bibitem{RogersSV} Rogers C., Saccomandi G. and Vergori L., Ermakov-modulated nonlinear Schr\"odinger models. Integrable reduction., J.
Nonlinear Math. Phys., 23 (2016) 106-126




\end{thebibliography}
\end{document}